\documentclass[11pt]{article}

\usepackage[numbers,square,sort&compress]{natbib}
\let\NumberedBibliography\thebibliography
\let\EndNumberedBibliography\endthebibliography
\usepackage[final]{acl}
\let\thebibliography\NumberedBibliography
\let\endthebibliography\EndNumberedBibliography
\setcitestyle{numbers,square,sort&compress}

\makeatletter
\renewcommand{\NAT@bibsetnum}[1]{%
  \setlength{\leftmargin}{\bibhang}%
  \setlength{\itemindent}{-\leftmargin}%
  \setlength{\labelwidth}{0pt}%
  \setlength{\labelsep}{0pt}%
  \setlength{\listparindent}{0pt}%
  \setlength{\itemsep}{\bibsep}%
  \setlength{\parsep}{0pt}%
}
\makeatother

\let\citeyearpar\citep

\usepackage{times}
\usepackage{latexsym}
\usepackage{microtype}
\usepackage{inconsolata}
\usepackage{amsmath,amssymb}
\usepackage{graphicx}
\usepackage{booktabs}

\providecommand{\Description}[1]{}
\hypersetup{hidelinks}
\usepackage[T1]{fontenc}

\usepackage[utf8]{inputenc}

\usepackage{multirow, makecell}
\usepackage{pifont}
\newcommand{\cmark}{\ding{51}}

\usepackage{tikz}
\usepackage{forest}
\usetikzlibrary{trees,positioning,shapes,shadows,arrows.meta}

\definecolor{ngreen}{HTML}{D5E8D4}
\definecolor{evalblue}{HTML}{D6E0F0}
\newcommand{\OoneM}{\textcolor{red}{\underline{O1}}}      
\newcommand{\OtwoM}{\textcolor{blue}{\underline{O2}}}     
\newcommand{\OthreeM}{\textcolor{green!60!black}{\underline{O3}}} 
\newcommand{\OtwoP}{\textcolor{blue}{O2}}

\newcommand{\TtwoP}{{T2}}               
\newcommand{\SecTag}[1]{\textcolor{gray}{sec:#1}}         
\newcommand{\PrimaryA}{\textcolor{gray}{\scriptsize primary:A}}
\newcommand{\PrimaryD}{\textcolor{gray}{\scriptsize primary:D}}
\newcommand{\PrimaryE}{\textcolor{gray}{\scriptsize primary:E}}

\usepackage{array}
\usepackage{tabularx}
\usepackage{enumitem}
\usepackage{adjustbox}
\usepackage{placeins}
\usepackage{balance}
\usepackage{eso-pic}

\newcommand{\AcceptanceNotice}{Accepted to \textit{Findings of the Association for Computational Linguistics: EMNLP 2026}.}

\title{A Survey of Secure Retrieval-Augmented Generation}
\newcommand{\AuthorNameFont}{\normalfont\fontsize{11pt}{13pt}\selectfont}
\author{
\AuthorNameFont Yuming Xu\textsuperscript{1,*}, Mingtao Zhang\textsuperscript{1,*}, Zhuohan Ge\textsuperscript{1,*}, Haoyang Li\textsuperscript{1,$\dagger$}, Nicole Hu\textsuperscript{1},\\
\AuthorNameFont Yongqi Zhang\textsuperscript{2}, Zhiyuan Wen\textsuperscript{1}, Jason Chen Zhang\textsuperscript{1}, Qing Li\textsuperscript{1}, Lei Chen\textsuperscript{2}\\
\normalfont\normalsize\textsuperscript{1}The Hong Kong Polytechnic University\\
\normalfont\normalsize\textsuperscript{2}The Hong Kong University of Science and Technology (Guangzhou)\\
\normalfont\normalsize\texttt{martin.xu@connect.polyu.hk}
}

\begin{document}
\AddToShipoutPictureFG*{%
  \AtPageLowerLeft{%
    \put(\LenToUnit{0.5\paperwidth},\LenToUnit{13mm}){%
      \makebox(0,0)[c]{\normalfont\fontsize{9pt}{11pt}\selectfont\AcceptanceNotice}%
    }%
  }%
}
\maketitle

\begin{abstract}

Retrieval-augmented generation (RAG) improves large language models (LLMs) with external knowledge, but this access path creates security risks distinct from inherent prompt-only or parametric-model flaws. We frame secure RAG as securing external knowledge access. We conducted a systematic search and curated 135 works on attacks, defenses, and security evaluation, and organized them with SLOT: a taxonomy along the attack Surface (S) and the corresponding defense Layer (L), with cross-cutting axes Objective (O) and attack-target scope (T). Mapping these studies onto an external knowledge-access pipeline, we expose three mismatches: target mismatch (T2 attacks outpace T2 defenses and evaluation), stage mismatch (S1 attacks outnumber L1 defenses), and signal mismatch (fluent, retrievable, corpus-fitting S1 attacks challenge anomaly-based L2/L3 defenses). Finally, we discuss directions for more realistic targets, surface-complete defense stacks, standardized evaluation, confidentiality, and multimodal and agentic systems, and release the screening process, selected-paper metadata, and machine-readable SLOT labels at \url{https://github.com/TreeAI-Lab/Awesome-RAG-Security}.

\end{abstract}
\hypersetup{pdfauthor={Yuming Xu, Mingtao Zhang, Zhuohan Ge, Haoyang Li, Nicole Hu, Yongqi Zhang, Zhiyuan Wen, Jason Chen Zhang, Qing Li, Lei Chen}}
\thispagestyle{empty}
\pagestyle{empty}
\begingroup
\renewcommand{\thefootnote}{}
\footnotetext{\textsuperscript{*}Denotes equal contribution.\quad \textsuperscript{$\dagger$}Corresponding author.}
\endgroup

\section{Introduction} \label{sec:intro}
Retrieval-augmented generation (RAG) has become a practical and widely adopted paradigm for improving large language models (LLMs) with external knowledge at inference time~\cite{lewis2020retrieval,gao2023retrieval,wu2024retrieval,gupta2024comprehensive,cheng2025survey,wang2024corag,zhu2026skyline}. 
By retrieving evidence from external corpora, databases, or structured repositories and incorporating it into the generation context, RAG improves factuality, {updatability}, and domain adaptability~\cite{gao2023retrieval,wu2024retrieval,yu2024evaluation,gan2025retrieval} and can make answers traceable to source evidence~\cite{liu2026polyuquest}. 
At the same time, this shift from parametric knowledge alone to external knowledge access introduces security risks, because the model is no longer affected only by its parameters and the user prompt, but also by {the content and access path of external knowledge}~\cite{ward2025adversarial,arzanipour2025rag,ammann2025securing}.

\begin{table*}[htbp]
	\caption{Scope comparison with adjacent RAG-centered surveys. EK Scope: external knowledge as a security object; Risk Bnd.: explicit RAG/generic-LLM boundary; Pipe. View: mapping to knowledge-access stages; A-D-E: attacks, defenses, and evaluation under one threat scope; Eval.: metrics, benchmarks, or protocols; Deploy.: mitigation, recovery, or access control. \protect\cmark/$\circ$/-- denote central, partial, and not-primary coverage.}
	\label{tab:survey_comparison}
	\centering
	\footnotesize
	\setlength{\tabcolsep}{3pt}
	\renewcommand{\arraystretch}{1.4}
	\resizebox{\textwidth}{!}{
		\begin{tabular}{@{}l >{\raggedright\arraybackslash}p{5.8cm} cccccc@{}}
			\toprule
			\textbf{Survey} & \textbf{Primary focus} & \textbf{EK Scope} & \textbf{Risk Bnd.} & \textbf{Pipe. View} & \textbf{A-D-E} & \textbf{Eval.} & \textbf{Deploy.} \\
			\midrule
			\citet{gao2023retrieval}       & RAG architectures and components                              & --      & --      & $\circ$ & --      & $\circ$ & --      \\
			\citet{yu2024evaluation}       & Evaluation metrics and benchmarks                             & --      & --      & $\circ$ & --      & \cmark  & --      \\
			\citet{wu2024retrieval}        & Retrieval modules and applications                            & --      & --      & $\circ$ & --      & $\circ$ & --      \\
			\citet{zhou2024trustworthiness}& Trustworthiness dimensions and benchmark                      & $\circ$ & $\circ$ & $\circ$ & $\circ$ & \cmark  & $\circ$ \\
			\citet{gupta2024comprehensive} & RAG landscape and future directions                           & --      & --      & $\circ$ & --      & --      & $\circ$ \\
			\citet{ni2025towards}          & Trustworthy RAG roadmap                                       & $\circ$ & $\circ$ & $\circ$ & $\circ$ & $\circ$ & $\circ$ \\
			\citet{cheng2025survey}        & Knowledge-oriented RAG taxonomy                               & $\circ$ & --      & $\circ$ & --      & $\circ$ & --      \\
			\citet{gan2025retrieval}       & Evaluation datasets and meta-analysis                         & --      & --      & $\circ$ & --      & \cmark  & --      \\
			\citet{ammann2025securing}     & Risk assessment and mitigation mapping                        & \cmark  & $\circ$ & \cmark  & $\circ$ & $\circ$ & \cmark  \\
			\citet{sharma2025retrieval}    & RAG robustness and benchmarks                                 & --      & --      & $\circ$ & --      & $\circ$ & $\circ$ \\
			\citet{ward2025adversarial}    & Adversarial threats and risk controls                         & \cmark  & $\circ$ & \cmark  & $\circ$ & --      & \cmark  \\
			\citet{arzanipour2025rag}      & RAG threat model and attack surface                           & \cmark  & \cmark  & \cmark  & $\circ$ & --      & --      \\
			\citet{bodea2026sok}           & Privacy risks and mitigations                                 & \cmark  & $\circ$ & \cmark  & $\circ$ & \cmark  & $\circ$ \\
			\midrule
			\textbf{Ours}                  & \textbf{External-knowledge security in RAG}                    & \cmark  & \cmark  & \cmark  & \cmark  & \cmark  & \cmark  \\
			\bottomrule
		\end{tabular}
	}

\end{table*}

\begin{figure}[t]
    \centering
    \includegraphics[width=\columnwidth]{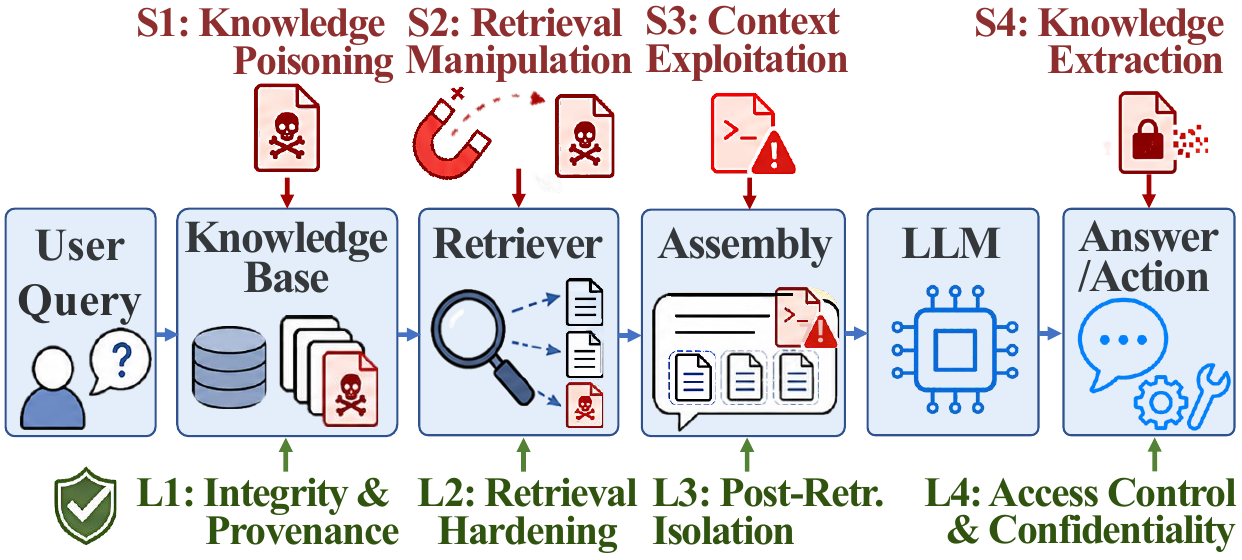}
    \Description{Four attack surfaces and their mirrored defense layers along a retrieval-augmented generation pipeline.}
    \caption{Attack surfaces ({S1--S4}) and defense layers that mirror them (L1--L4) along the RAG pipeline.}
    \label{fig:rag_security}
\end{figure}

Research on RAG security is growing rapidly, yet individual studies usually attack or defend one setting without a shared problem definition {or taxonomy} to relate these efforts. 
Recent surveys cover general architectures and benchmarks~\cite{gao2023retrieval,yu2024evaluation,gan2025retrieval}, trustworthiness, adversarial robustness, and privacy~\cite{zhou2024trustworthiness,ni2025towards,ammann2025securing,ward2025adversarial,arzanipour2025rag,bodea2026sok}. 
Table~\ref{tab:survey_comparison} compares their scope. To our knowledge, no prior survey jointly analyzes RAG security through the external-knowledge-access pipeline, distinguishes risks introduced or amplified by RAG from risks inherent to LLMs, and links attacks, defenses, and benchmarks. 
We therefore present the first systematic, external-knowledge-centered survey dedicated to RAG security under one hierarchical taxonomy.

As shown in Figure~\ref{fig:rag_security}, the central problem is that an attacker need not touch the model or the user-visible prompt. {By manipulating external content or retrieval or delivering instructions through retrieved context, an attacker can steer generation or extract protected knowledge.}
Each point where attackers can act mirrors a point where defenders can intervene, which motivates our \emph{Surface$\leftrightarrow$Layer} mirroring as the backbone of the survey.

To keep this backbone focused, we define an \emph{operational boundary}: a risk is in scope only if external knowledge carries the threat, creates an entry point absent in prompt-only LLM use, or materially amplifies its persistence, transferability, or blast radius. 
Generic prompt-only jailbreaks and purely parametric memorization are therefore excluded. 
Under this boundary, we organize attack-side work by the attacked surface (S), mirror each surface with the defense layer (L) that controls the same point, 
and tag every method with its {attack-target scope (T)} and security objective (O). 
We then discuss benchmark and evaluation studies on this attack-defense correspondence and distill insights and concrete future directions.

Concretely, our contributions are:
\begin{itemize}[leftmargin=*]
    \item \textbf{Scope and protocol.} We define secure RAG around risks introduced or amplified by external knowledge access, and use a systematic protocol covering May 1, 2020--April 30, 2026.

    \item \textbf{The SLOT taxonomy.} We organize RAG security by attack \emph{Surface} (S) and corresponding defense \emph{Layer} (L), then cross-tag security \emph{Objective} (O) and {attack-target scope} (T).

    \item \textbf{A {135-study} synthesis.} We jointly analyze {64 attack-primary, 56 defense-primary, and 15 evaluation-primary works}, connecting attacks, defenses, and evaluation along the pipeline.

    \item \textbf{A quantitative diagnosis of mismatches and future directions.} We identify {three mismatches: a target mismatch (31/64 T2 attacks versus 2/56 T2 defenses and 1/15 T2-centered evaluation studies), a stage mismatch (42 primary-S1 attacks versus 4 primary-L1 defenses), and a signal mismatch (fluent, retrievable, corpus-fitting S1 attacks versus L2/L3 defenses that often assume anomalous or separable poison)}. We then discuss directions for attacks, defenses, and benchmarks.

    \item \textbf{An open, machine-readable community resource.} We release {the screening process, selected-paper metadata, and machine-readable SLOT labels for reuse} in the \href{https://github.com/TreeAI-Lab/Awesome-RAG-Security}{Awesome RAG Security repository}, helping the community track new studies, align them under a shared taxonomy, and identify gaps and research opportunities.
\end{itemize}

\section{RAG Pipeline and Operational Scope}\label{sec:bg}

\begin{figure*}[t]
    \centering
    \includegraphics[width=0.99\textwidth]{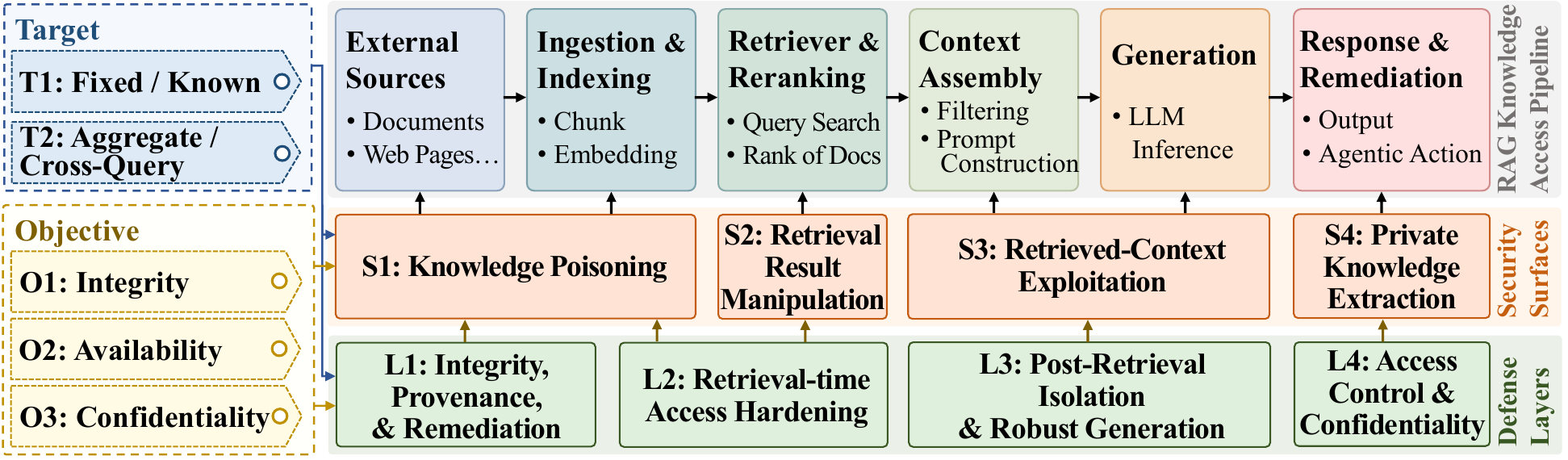}
    \Description{The six stages of external knowledge access, aligned with SLOT surfaces, defense layers, objectives, and target scopes.}
    \caption{
        SLOT view of the RAG knowledge-access pipeline. 
        Attack surfaces ({S1--S4}) and defense layers ({L1--L4}) are aligned with the pipeline stages, 
        while Objective (O) and Target (T) are cross-cutting tags used to compare attacks, defenses, and benchmarks, 
        i.e., SLOT Tag = Surface/Layer + Objective + Target.
    }
    \label{fig:pipeline}
\end{figure*}

\subsection{External Knowledge Access Pipeline with Security Surfaces}\label{sec:security_surfaces}

As shown in Figure~\ref{fig:pipeline}, we view RAG-related processes through the {external-knowledge-access pipeline}, 
which can be divided into six stages: 
external sources with raw content; 
ingestion and indexing that turn raw content into chunks and embeddings; 
retrieval and reranking select candidate evidence; 
context assembly formats the selected evidence into the model-visible prompt; 
generation produces the answer; 
and response and remediation deliver the final output. 
We refer to the indexed external corpus, database, memory, or graph-based store~\cite{wang2026gorag,wang2026agrag} as the \emph{knowledge substrate}, and to the inference-time selection and use of its evidence as \emph{external knowledge access}.
\emph{Protected knowledge} is substrate content governed by a disclosure policy: it may support an answer but must not be revealed beyond what that policy permits. 
\emph{Disclosure} is the release or reconstruction of protected knowledge beyond those limits.

The pipeline also identifies four attack surfaces: pre-retrieval knowledge poisoning (S1), retrieval-time access manipulation (S2), post-retrieval context exploitation (S3), and knowledge exfiltration and privacy attacks (S4). Defenses mirror them through knowledge-base integrity, provenance, and remediation (L1); retrieval-time access hardening (L2); post-retrieval isolation and robust generation (L3); and access control and confidentiality (L4).

\subsection{Threat Model}\label{ssec:threat}
In our discussion, an attack acts through or exploits the external-knowledge-access path. 
An attacker may influence ingestible or retrievable content, manipulate retrieval or its components, deliver instructions through retrieved context, or issue probing queries to infer protected knowledge. 
A compromised retriever or generator is included only when the threat fundamentally depends on external knowledge access. 
Attacks confined to the user prompt or model parameters without using this path are outside our scope. 
The surface (S1--S4) identifies where the attacker acts, and the mirrored layer (L1--L4) identifies where a defense intervenes along the same pipeline.

The objective (O) is the security property that the attack seeks to violate, following the classical CIA triad~\cite{saltzer1975protection}. 
Integrity (O1) attacks push the system to produce, retrieve, or act on attacker-chosen content; 
availability (O2) attacks make it refuse or fail to answer legitimate queries; 
and confidentiality (O3) attacks infer or extract protected knowledge from the substrate. 
Any objective can escalate from an incorrect or leaked answer into an unsafe action.

The target axis (T) describes what the attacker seeks to affect and at what scope. 
T1 (\emph{fixed/known target}) assumes that a finite query set is fixed and known to the attacker before the attack, while T2 (\emph{aggregate/cross-query target}) seeks an aggregate effect over unknown future queries or adaptively explored records. 
For O1/O2, T1 changes answers or denies service to known queries, while T2 steers answers or causes persistent refusal across a future query family about an attacker-specified entity, topic, or claim.
For O3, T1 infers or extracts a pre-specified record, while T2 maximizes recovery of initially unknown corpus content through repeated or adaptive queries. 
T2 therefore requires an aggregate view over a query distribution rather than a per-instance check.

Together, these dimensions define SLOT. 
A threat receives S/O/T tags, and a defense receives L/O/T tags. 
A defense or evaluation inherits T from the attack-target setting that it protects against or measures.
Primary S/L and O mark the main intervention point and objective, while secondary labels record additional roles. 
Counts use only the primary paper type and S/L label, while works under a non-primary category are excluded and reported in gray.
For example, \emph{PoisonedRAG}~\cite{zou2025poisonedrag} is S1/O1/T1 as it injects passages before retrieval to alter answers to known queries; \emph{RobustRAG}~\cite{xiang2024certifiably} is L3/O1/T1 because it processes retrieved passages and aggregates outputs under a fixed-query corruption bound. 
Table~\ref{tab:slot_examples} in Section~\ref{ssec:slot_coding} gives examples for all S/L categories.

\subsection{SLOT Coding Guide}

\label{ssec:slot_coding}

Primary S/L identifies the first pipeline point at which a paper's central attack or control acts; a secondary S/L records a substantial mechanism at another stage. Primary O identifies the main measured security objective, while secondary objectives record additional measured effects. {T1 fixes the target instance, query, or finite query set and assumes that the attacker knows it before the attack; T2 targets aggregate behavior over unknown future queries or adaptively explored records.} {Each work is counted once as attack-, defense-, or evaluation-primary; secondary roles are retained and shown in gray when cross-listed.}

\begin{table*}[t]
    \caption{Representative examples for the eight primary Surface/Layer categories. Secondary labels are retained in the released records.}
    \label{tab:slot_examples}
    \centering
    \small
    \setlength{\tabcolsep}{5pt}
    \renewcommand{\arraystretch}{1.15}
    \begin{tabularx}{\textwidth}{@{}
        >{\centering\arraybackslash}p{0.10\textwidth}
        >{\raggedright\arraybackslash}p{0.245\textwidth}
        >{\centering\arraybackslash}p{0.095\textwidth}
        >{\raggedright\arraybackslash}X@{}}
        \toprule
        \textbf{Category} & \textbf{Representative work} & \textbf{O/T} & \textbf{Reason for primary placement} \\
        \midrule
        S1 & PoisonedRAG~\cite{zou2025poisonedrag} & O1/T1 & Injects malicious knowledge before retrieval. \\
        {S2} & {GGPP}~\cite{hu2024ggpp} & {O1/T1} & {Manipulates a known query to retrieve false evidence.} \\
        {S3} & {Indirect Injection}~\cite{greshake2023not} & {O1/T2} & {Delivers instructions through retrieved external content.} \\
        S4 & MIA-RAG~\cite{anderson2025my} & O3/T1 & Infers membership in a private corpus. \\
        {L1} & {D-RAG}~\cite{e_andersen2025d} & {O1/T1} & {Applies provenance controls to the knowledge substrate.} \\
        {L2} & {RAGuard-Defense}~\cite{cheng2025secure} & {O1/T1} & {Filters retrieved candidates before generation.} \\
        L3 & InstructRAG~\cite{weiinstructrag} & O1/T1 & Limits malicious post-retrieval influence. \\
        {L4} & {AC-RAG}~\cite{10.1145/3672608.3707848} & {O3/T1} & {Enforces fine-grained authorization for retrieved knowledge.} \\
        \bottomrule
    \end{tabularx}

\end{table*}

\subsection{Operational Scope}\label{ssec:scope}
We include a work when external knowledge carries the threat, creates an attack path absent from prompt-only LLM use, or materially amplifies the threat through persistence, transferability, or broader exposure.
Thus, Greshake et al.~\cite{greshake2023not} and BIPIA~\cite{yi2025benchmarking} are included because retrieved external content carries indirect instructions, while JailbreakBench~\cite{chao2024jailbreakbench} is excluded because it evaluates prompt-only attacks, and RAGBench~\cite{friel2024ragbench} is excluded because it evaluates ordinary RAG quality. 
We also exclude purely parametric memorization, generic LLM safety or privacy, and work that only uses RAG for unrelated applications.

\noindent\textbf{Relation to general LLM security.}
General LLM defenses help at the stages they protect. Prompt-injection detection or model alignment can reduce harmful behavior after retrieved content reaches the generator, but neither by itself prevents corpus poisoning, retrieval manipulation, or private-knowledge extraction. Secure RAG therefore also needs controls on the external-knowledge path, such as provenance and admission checks, retrieval hardening, and access control.

\subsection{Literature Search Protocol}\label{ssec:search_protocol}
Following the operational scope above, we conducted a structured search of five scholarly sources.

\noindent\textbf{Sources.} We searched arXiv, ACL Anthology, OpenReview, Semantic Scholar, and DBLP.

\noindent\textbf{Date range.} We retained works first made public from May 1, 2020 to April 30, 2026.

\noindent\textbf{Query rule.} We required each record to contain at least one \emph{RAG} term and one \emph{security} term in its title or abstract. Terms are reported in Appendix~\ref{app:search_protocol}.

\noindent\textbf{Screening and selection.}
Source retrieval: 17,256 records. Abstract-available deduplication: 8,262. Title/abstract screening with \texttt{DeepSeek-V4-}\allowbreak\texttt{Flash-0731}: 636. Version merging: 617. 
Full-text screening with the same model and scope rule: 552 eligible works. 
Expert selection: 132. 
Snowballing: $+3$, yielding 135 studies. 
Appendix~\ref{app:search_protocol} reports the prompt templates and detailed screening rules, with the full funnel in Table~\ref{tab:screening_funnel}; the \href{https://github.com/TreeAI-Lab/Awesome-RAG-Security}{repository} releases the screening process, selected-paper metadata with public paper and code links, and machine-readable SLOT labels.

\begin{table}[tbp]
    \caption{Literature-search and screening funnel.}
    \label{tab:screening_funnel}
    \centering
    \setlength{\tabcolsep}{8pt}
    \begin{tabular}{@{}p{0.68\columnwidth}>{\raggedleft\arraybackslash}p{0.24\columnwidth}@{}}
        \toprule
        \textbf{Stage} & \textbf{Retained} \\
        \midrule
        Source retrieval & 17,256 \\
        Source consolidation & 8,262 \\
        Title/abstract inclusion & 636 \\
        Version-level unique works & 617 \\
        Full-text eligibility & 552 \\
        Expert selection & 132 \\
        Snowballing & +3 \\
        \textbf{Final corpus} & \textbf{135} \\
        \bottomrule
    \end{tabular}

\end{table}

\section{Attack Mechanisms}\label{sec:attack}
Following Section~\ref{ssec:threat}, we organize attacks by their first intervention surface (S1--S4), objective (O1--O3), and target scope (T1--T2). Methods with the same primary mechanism are grouped below; Figure~\ref{fig:taxonomy_attack} gives the full taxonomy.

\begin{figure*}[tp]
\centering
\small
\tikzset{
    basic/.style  = {draw, text width=2cm, align=center, font=\sffamily, rectangle},
    root/.style   = {basic, rounded corners=2pt, thin, fill=white, text width=7.2cm, rotate=90, font=\footnotesize},
    mnode/.style  = {basic, thin, rounded corners=2pt, fill=ngreen, text width=2.60cm, font=\footnotesize},
    mnode_1/.style = {basic, thin, rounded corners=2pt, fill=ngreen, text width=2.25cm, font=\footnotesize},
    wnode/.style  = {basic, thin, rounded corners=2pt, align=left, fill=white, text width=9.15cm, font=\footnotesize},
}
\begin{adjustbox}{max width=\textwidth,max totalheight=0.82\textheight}
\begin{forest}
for tree={
    grow=east, reversed, growth parent anchor=east, parent anchor=east, child anchor=west,
    l sep=0.20cm, s sep=0.55mm,
    edge path={\noexpand\path[\forestoption{edge},->, >={latex}]
         (!u.parent anchor) -- +(2pt,0pt) |- (.child anchor)
         \forestoption{edge label};}
}
[RAG Attack and Audit Methods, root, parent anchor=south
    [Knowledge\\Poisoning (S1), mnode_1
        [Query-Specific Corpus\\Poisoning, mnode
            [{\renewcommand{\arraystretch}{0.94}\begin{tabular}{@{}p{4.44cm}@{\hspace{0.08cm}}p{4.44cm}@{}}
            PoisonedRAG~\cite{zou2025poisonedrag}\textsuperscript{\SecTag{S2},\OoneM} & GARAG~\cite{cho2024garag}\textsuperscript{\SecTag{S2},\OoneM}\\
            Retrieval Poisoning~\cite{zhang2024retrievalpoisoning}\textsuperscript{\SecTag{S3},\OoneM} & Silent Saboteur~\cite{song2025regent}\textsuperscript{\SecTag{S2},\OoneM}\\
            CorruptRAG~\cite{zhang2025practical}\textsuperscript{\SecTag{S2},\OoneM} & AuthChain~\cite{chang2025authchain}\textsuperscript{\SecTag{S2},\OoneM}\\
            HijackRAG~\cite{zhang2024hijackrag}\textsuperscript{\SecTag{S2/S3},\OoneM} & RIPRAG~\cite{xi2026riprag}\textsuperscript{\SecTag{S2},\OoneM}\\
            RAG Paradox~\cite{choi2025ragparadox}\textsuperscript{\SecTag{S2},\OoneM} & Token-Level Attack~\cite{li2026tokenlevel}\textsuperscript{\SecTag{S2},\OoneM}\\
            Joint-GCG~\cite{wang2026jointgcg}\textsuperscript{\SecTag{S2/S3},\OoneM} & 
            \end{tabular}}, wnode]
        ]
        [Trigger / Generalized\\Poisoning, mnode
            [{\renewcommand{\arraystretch}{0.94}\begin{tabular}{@{}p{4.44cm}@{\hspace{0.08cm}}p{4.44cm}@{}}
            {BadRAG}~\cite{xue2024badrag}\textsuperscript{\SecTag{S3},\OoneM,\OtwoP,\TtwoP} & GASLITE~\cite{bentov2025gaslite}\textsuperscript{\SecTag{S2},\OoneM,\TtwoP}\\
            UniC-RAG~\cite{geng2025unicrag}\textsuperscript{\SecTag{S2/S3},\OoneM,\TtwoP} & MIRAGE~\cite{chen2026miragemisleadingretrievalaugmentedgeneration}\textsuperscript{\SecTag{S2/S3},\OoneM,\TtwoP}\\
            AIP~\cite{chaturvedi2025aip}\textsuperscript{\SecTag{S2},\OoneM,\TtwoP} & AGGD~\cite{su2025corpuspoisoning}\textsuperscript{\SecTag{S2},\OoneM,\TtwoP}\\
            LIAR~\cite{tan2024gluepizza}\textsuperscript{\SecTag{S2/S3},\OoneM,\TtwoP} & Influential Tokens~\cite{wang2025influentialtokens}\textsuperscript{\SecTag{S2/S3},\OoneM,\OtwoM,\TtwoP}\\
            EmoRAG~\cite{zhou2026emorag}\textsuperscript{\SecTag{S2/L2},\OoneM,\TtwoP} & FlippedRAG~\cite{chen2024black}\textsuperscript{\SecTag{S2},\OoneM}\\
            Topic-FlipRAG~\cite{gong2025topicfliprag}\textsuperscript{\SecTag{S2},\OoneM,\TtwoP} & Backdoored Retrievers~\cite{clop2024backdooredretrievers}\textsuperscript{\SecTag{S3},\OoneM,\TtwoP}\\
            PR-Attack~\cite{jiao2025prattack}\textsuperscript{\SecTag{S2/S3},\OoneM} & PANDORA~\cite{deng2024pandora}\textsuperscript{\SecTag{S2/S3},\OoneM,\TtwoP}\\
            Phantom~\cite{chaudhari2024phantom}\textsuperscript{\SecTag{S2/S3},\OoneM,\OtwoM,\OthreeM,\TtwoP} & ConfusedPilot~\cite{roychowdhury2024confusedpilot}\textsuperscript{\SecTag{S2/S3/S4},\OoneM,\OthreeM,\TtwoP}\\
            PIDP-Attack~\cite{wang2026pidpattack}\textsuperscript{\SecTag{S2},\OoneM,\TtwoP} & Blocker~\cite{shafran2025blockerdocuments}\textsuperscript{\SecTag{S2/S3},\OtwoM}\\
            MutedRAG~\cite{suo2025mutedrag}\textsuperscript{\SecTag{S2/S3},\OtwoM} & 
            \end{tabular}}, wnode]
        ]
        [Agent Memory\\Poisoning, mnode
            [{\renewcommand{\arraystretch}{0.94}\begin{tabular}{@{}p{4.44cm}@{\hspace{0.08cm}}p{4.44cm}@{}}
            AgentPoison~\cite{chen2024agentpoisonredteamingllmagents}\textsuperscript{\SecTag{S3},\OoneM,\TtwoP} & MemoryGraft~\cite{srivastava2025memorygraftpersistentcompromisellm}\textsuperscript{\SecTag{S3},\OoneM,\TtwoP}\\
            MINJA~\cite{dong2026memoryinjectionattacksllm}\textsuperscript{\SecTag{S3},\OoneM,\TtwoP} & 
            \end{tabular}}, wnode]
        ]
        [Data-Loading\\Poisoning, mnode
            [{\renewcommand{\arraystretch}{0.94}\begin{tabular}{@{}p{4.44cm}@{\hspace{0.08cm}}p{4.44cm}@{}}
            {PhantomText Toolkit}~\cite{castagnaro2025dataloaders}\textsuperscript{\OoneM,\TtwoP} &
            \end{tabular}}, wnode]
        ]
        [Graph / Multimodal\\Knowledge Poisoning, mnode
            [{\renewcommand{\arraystretch}{0.94}\begin{tabular}{@{}p{4.44cm}@{\hspace{0.08cm}}p{4.44cm}@{}}
            GRAGPoison~\cite{liang2025gragpoison}\textsuperscript{\SecTag{S2},\OoneM} & {TKPA/UKPA}~\cite{wen2025tkpaukpa}\textsuperscript{\OoneM,\OtwoP}\\
            LogicPoison~\cite{xiao2026logicpoison}\textsuperscript{\OoneM} & MM-PoisonRAG~\cite{ha2025mmpoisonrag}\textsuperscript{\SecTag{S2/S3},\OoneM,\OtwoM}\\
            Poisoned-MRAG~\cite{liu2025poisonedmrag}\textsuperscript{\SecTag{S2},\OoneM} & One Pic~\cite{shereen2026onepic}\textsuperscript{\SecTag{S2/S3},\OoneM,\OtwoM,\TtwoP}
            \end{tabular}}, wnode]
        ]
        [Code-Oriented\\Knowledge Poisoning, mnode
            [{\renewcommand{\arraystretch}{0.94}\begin{tabular}{@{}p{4.44cm}@{\hspace{0.08cm}}p{4.44cm}@{}}
            RACG~\cite{lin2025kbpoisoning}\textsuperscript{\OoneM} & ImportSnare~\cite{ye2025importsnare}\textsuperscript{\SecTag{S2/S3},\OoneM,\TtwoP}
            \end{tabular}}, wnode]
        ]
    ]
    [Retrieval Result\\Manipulation (S2), mnode_1
        [Query / Retriever\\Manipulation, mnode
            [{\renewcommand{\arraystretch}{0.94}\begin{tabular}{@{}p{4.44cm}@{\hspace{0.08cm}}p{4.44cm}@{}}
            GGPP~\cite{hu2024ggpp}\textsuperscript{\SecTag{L2},\OoneM} & Medusa~\cite{shang2025medusacrossmodaltransferableadversarial}\textsuperscript{\OoneM}\\
            RAG-Pull~\cite{stambolic2025ragpull}\textsuperscript{\SecTag{S1},\OoneM,\TtwoP} & TrojanRAG~\cite{cheng2024trojanrag}\textsuperscript{\SecTag{S1/S3},\OoneM,\TtwoP}\\
            Your RAG is Unfair~\cite{bagwe2025unfair}\textsuperscript{\SecTag{S1},\OoneM,\TtwoP} & 
            \end{tabular}}, wnode]
        ]
    ]
    [Retrieved-Context\\Exploitation (S3), mnode_1
        [Indirect Prompt\\Injection, mnode
            [{\renewcommand{\arraystretch}{0.94}\begin{tabular}{@{}p{4.44cm}@{\hspace{0.08cm}}p{4.44cm}@{}}
            Indirect Injection~\cite{greshake2023not}\textsuperscript{\SecTag{S1},\OoneM,\TtwoP} & \textcolor{gray}{BIPIA~\cite{yi2025benchmarking}}\textsuperscript{\PrimaryE,\SecTag{L3},\OoneM}
            \end{tabular}}, wnode]
        ]
    ]
    [Private Knowledge\\Extraction (S4), mnode_1
        [Membership Inference\\and Auditing, mnode
            [{\renewcommand{\arraystretch}{0.94}\begin{tabular}{@{}p{4.44cm}@{\hspace{0.08cm}}p{4.44cm}@{}}
            MIA-RAG~\cite{anderson2025my}\textsuperscript{\SecTag{L3},\OthreeM} & S$^2$MIA~\cite{li2025generating}\textsuperscript{\OthreeM}\\
            MBA~\cite{liu2025mask}\textsuperscript{\OthreeM} & IA~\cite{naseh2025riddle}\textsuperscript{\OthreeM}\\
            S-RAG~\cite{zeng2025s}\textsuperscript{\OthreeM} & SMA~\cite{sun2025sma}\textsuperscript{\OthreeM}
            \end{tabular}}, wnode]
        ]
        [Record / Multimodal\\Extraction, mnode
            [{\renewcommand{\arraystretch}{0.94}\begin{tabular}{@{}p{4.44cm}@{\hspace{0.08cm}}p{4.44cm}@{}}
            Good and Bad~\cite{zeng2024good}\textsuperscript{\SecTag{S2/S3/L2/L3},\OthreeM} & Beyond Text~\cite{zhang2025beyond}\textsuperscript{\SecTag{S2/S3},\OthreeM}\\
            {mRAG Privacy}~\cite{al2026systemic}\textsuperscript{\OthreeM} & Backdoor Extraction~\cite{peng2024backdoorextraction}\textsuperscript{\OthreeM}
            \end{tabular}}, wnode]
        ]
        [Aggregate / Adaptive\\Extraction, mnode
            [{\renewcommand{\arraystretch}{0.94}\begin{tabular}{@{}p{4.44cm}@{\hspace{0.08cm}}p{4.44cm}@{}}
            Spill the Beans~\cite{qi2025spillthebeans}\textsuperscript{\SecTag{S3/L3},\OthreeM,\TtwoP} & IKEA~\cite{wang2025ikea}\textsuperscript{\OthreeM,\TtwoP}\\
            SECRET~\cite{he2025external}\textsuperscript{\SecTag{S2/S3},\OthreeM,\TtwoP} & RAGCRAWLER~\cite{yao2026ragcrawler}\textsuperscript{\OthreeM,\TtwoP}
            \end{tabular}}, wnode]
        ]
        [Graph Extraction, mnode
            [{\renewcommand{\arraystretch}{0.94}\begin{tabular}{@{}p{4.44cm}@{\hspace{0.08cm}}p{4.44cm}@{}}
            GraphRAG Privacy~\cite{liu2026graphprivacy}\textsuperscript{\SecTag{S2/S3},\OthreeM,\TtwoP} & Agentic Graph Extraction~\cite{yang2026agenticgraph}\textsuperscript{\SecTag{S2/S3},\OthreeM,\TtwoP}\\
            \textcolor{gray}{Retrieval Pivot~\cite{thornton2026retrievalpivotattacks}}\textsuperscript{\PrimaryD,\OthreeM} &
            \end{tabular}}, wnode]
        ]
    ]
]
\end{forest}
\end{adjustbox}
\caption{
    Taxonomy of RAG attack methods by primary attack surface (S1--S4).
    {Each attack-primary method appears once at its first intervention point. 
    The gray BIPIA and Retrieval Pivot cross-lists are evaluation- and defense-primary, respectively, and excluded from attack counts.}
    \textcolor{gray}{Gray sec:} tags record explicit secondary surfaces or defense layers and are excluded from primary counts.
    Superscripts mark attack objectives: \textcolor{red}{O1} integrity, \textcolor{blue}{O2} availability, and \textcolor{green!60!black}{O3} confidentiality.
    {T2} denotes {aggregate targets over unknown future queries or adaptively explored records, marking 31 of 64 attack-primary works}; unmarked works are T1.
}
\label{fig:taxonomy_attack}
\Description{A hierarchical taxonomy of 64 attack-primary studies by surface and mechanism, with objective, target-scope, and secondary-role tags.}
\end{figure*}

\subsection{Knowledge Poisoning (S1)}\label{ssec:attack:s1}
Knowledge poisoning inserts malicious corpus, file, graph, or memory content before retrieval, allowing the payload to persist across queries or sessions. Most methods target integrity (O1), while some also induce refusal or failure (O2).

Fixed-query text attacks share a known query or answer target but differ in how they construct the poison. 
\emph{PoisonedRAG}~\cite{zou2025poisonedrag} makes injected passages both retrievable and answer-directing, while \emph{GARAG}~\cite{cho2024garag} and \emph{Retrieval Poisoning}~\cite{zhang2024retrievalpoisoning} emphasize natural, highly retrievable text. 
\emph{The Silent Saboteur}~\cite{song2025regent} edits relevant documents, while \emph{CorruptRAG}~\cite{zhang2025practical} and \emph{AuthChain}~\cite{chang2025authchain} embed coherent target evidence. 
\emph{HijackRAG}~\cite{zhang2024hijackrag} and \emph{Joint-GCG}~\cite{wang2026jointgcg} couple retrieval with downstream generation through prompt hijacking or joint optimization. 
\emph{RIPRAG}~\cite{xi2026riprag}, \emph{The RAG Paradox}~\cite{choi2025ragparadox}, and \emph{Token-Level Attack}~\cite{li2026tokenlevel} use reinforcement learning, live feedback, or token substitutions.

{Trigger-based and generalized attacks vary} in how broadly the payload transfers. 
Trigger-activated methods make poisoned evidence rank highly when an attacker-defined signal matches. \emph{BadRAG}~\cite{xue2024badrag} and \emph{AIP}~\cite{chaturvedi2025aip} associate natural semantic triggers with poisoned passages or reusable instructional prompts, \emph{Backdoored Retrievers}~\cite{clop2024backdooredretrievers} implants the association during retriever fine-tuning, and \emph{PR-Attack}~\cite{jiao2025prattack} jointly optimizes a soft-prompt trigger and poison texts. \emph{AGGD}~\cite{su2025corpuspoisoning} uses no explicit inference-time trigger; it instead searches for a universal adversarial passage that acts as an always-on retrieval lure across unseen queries and domains. Distribution-oriented methods likewise avoid assuming the exact future query, but optimize poison against a sampled query distribution: \emph{GASLITE}~\cite{bentov2025gaslite} uses embedding gradients to rank attacker content for unseen concept-related queries, \emph{UniC-RAG}~\cite{geng2025unicrag} clusters diverse target queries and jointly optimizes a small set of texts for retrieval and generation, and \emph{MIRAGE}~\cite{chen2026miragemisleadingretrievalaugmentedgeneration} approximates latent queries through persona-driven synthesis and embeds semantic anchors under black-box feedback. \emph{LIAR (Glue Pizza and Eat Rocks)}~\cite{tan2024gluepizza} and \emph{Influential Tokens}~\cite{wang2025influentialtokens} couple retrievability with misleading or refusal behavior, whereas \emph{EmoRAG}~\cite{zhou2026emorag}, \emph{FlippedRAG}~\cite{chen2024black}, and \emph{Topic-FlipRAG}~\cite{gong2025topicfliprag} steer affect or stance. {Payload-bearing variants use retrieved content downstream: \emph{PANDORA}~\cite{deng2024pandora} induces jailbreaks; \emph{Phantom}~\cite{chaudhari2024phantom} and \emph{ConfusedPilot}~\cite{roychowdhury2024confusedpilot} carry malicious commands or enterprise instructions; and \emph{Blocker}~\cite{shafran2025blockerdocuments} and \emph{MutedRAG}~\cite{suo2025mutedrag} induce refusal. \emph{PIDP-Attack}~\cite{wang2026pidpattack} instead combines poisoned passages with a query suffix (secondary S2).}

Other S1 methods adapt the same persistence mechanism to different substrates. 
In agent memory, \emph{AgentPoison}~\cite{chen2024agentpoisonredteamingllmagents}, \emph{MemoryGraft}~\cite{srivastava2025memorygraftpersistentcompromisellm}, and \emph{MINJA}~\cite{dong2026memoryinjectionattacksllm} plant records that affect later retrieval, planning, or actions. 
\emph{PhantomText Toolkit}~\cite{castagnaro2025dataloaders} instead hides payloads in the data-loading path.
In graph RAG, \emph{GRAGPoison}~\cite{liang2025gragpoison}, \emph{TKPA/UKPA}~\cite{wen2025tkpaukpa}, and \emph{LogicPoison}~\cite{xiao2026logicpoison} manipulate graph evidence or reasoning, with UKPA additionally inducing broad O2 degradation. 
\emph{MM-PoisonRAG}~\cite{ha2025mmpoisonrag}, \emph{Poisoned-MRAG}~\cite{liu2025poisonedmrag}, and \emph{One Pic}~\cite{shereen2026onepic} poison multimodal evidence, 
while \emph{RACG}~\cite{lin2025kbpoisoning} and \emph{ImportSnare}~\cite{ye2025importsnare} steer code generation through poisoned examples, documentation, or dependencies. 
The move toward fluent, retrievable, cross-query payloads creates the signal mismatch {discussed in} Sections~\ref{sec:defense} and~\ref{sec:insights}.

\subsection{Retrieval Result Manipulation (S2)}\label{ssec:attack:s2}
S2 attacks alter the query, ranking, or retriever so attacker-preferred evidence reaches the generator. 
Query-side methods modify the retrieval request, including \emph{GGPP}~\cite{hu2024ggpp} that appends optimized tokens to retrieve false documents, \emph{Medusa}~\cite{shang2025medusacrossmodaltransferableadversarial} that perturbs visual queries, and \emph{RAG-Pull}~\cite{stambolic2025ragpull} that shifts query embeddings toward attacker-chosen code gadgets across future queries. 
Retriever-side methods instead implant persistent, trigger-conditioned behavior. \emph{TrojanRAG}~\cite{cheng2024trojanrag} binds triggers to a compromised retriever and poisoned context, while \emph{Your RAG is Unfair}~\cite{bagwe2025unfair} backdoors a reusable encoder to induce demographic bias. GGPP is attack-primary, and its detector is a secondary defense contribution.

\subsection{Retrieved-Context Exploitation (S3)} \label{ssec:attack:s3}
S3 attacks use model-visible external content to carry instructions, jailbreaks, or refusal triggers after retrieval. 
\emph{Indirect Injection}~\cite{greshake2023not} establishes this attack path, while \emph{BIPIA}~\cite{yi2025benchmarking} systematizes its tasks, goals, injection positions, and defenses. 
BIPIA is evaluation-primary and included as a cross-list. 
Among S1-primary attacks, \emph{PANDORA}~\cite{deng2024pandora} uses retrieved harmful material for jailbreaks, \emph{Phantom}~\cite{chaudhari2024phantom} and \emph{ConfusedPilot}~\cite{roychowdhury2024confusedpilot} embed commands or enterprise instructions, while \emph{Blocker}~\cite{shafran2025blockerdocuments} and \emph{MutedRAG}~\cite{suo2025mutedrag} induce refusal. They are S1-primary with corpus insertion as their first intervention.

\subsection{Private Knowledge Extraction (S4)} \label{sec:attack_exfiltration}
S4 attacks infer or recover protected substrate content (O3). 
First, membership methods test a fixed candidate (T1). 
\emph{MIA-RAG}~\cite{anderson2025my} uses crafted prompts; \emph{S$^2$MIA}~\cite{li2025generating} compares target samples with generated text; \emph{MBA}~\cite{liu2025mask} masks unrelated-document interference; while \emph{IA}~\cite{naseh2025riddle} asks questions whose answers depend on document presence. 
\emph{S-RAG}~\cite{zeng2025s} and \emph{SMA}~\cite{sun2025sma} adapt membership inference for privacy auditing and source attribution.

Targeted extraction recovers selected records or modalities. 
\emph{The Good and the Bad}~\cite{zeng2024good} studies retrieval-induced leakage and memorization trade-offs; 
\emph{Beyond Text}~\cite{zhang2025beyond} extends extraction to vision- and speech-language RAG; 
\emph{mRAG Privacy}~\cite{al2026systemic} evaluates image membership and caption recovery; 
and \emph{Backdoor Extraction}~\cite{peng2024backdoorextraction} leaks retrieved references through a compromised generator. 
These works use pre-specified targets or evaluation cases and therefore are T1.

Aggregate extraction instead maximizes recovery of initially unknown content. 
\emph{Spill the Beans}~\cite{qi2025spillthebeans} induces datastore regurgitation, 
\emph{IKEA}~\cite{wang2025ikea} extracts implicit knowledge without explicit jailbreaks, 
\emph{SECRET}~\cite{he2025external} composes extraction instructions, jailbreak operators, and retrieval triggers, 
and \emph{RAGCRAWLER}~\cite{yao2026ragcrawler} adaptively follows corpus relations. 
\emph{Exposing Privacy Risks in GraphRAG}~\cite{liu2026graphprivacy} and \emph{Agentic Graph Extraction}~\cite{yang2026agenticgraph} extend aggregate recovery to graph neighborhoods. 
These attack methods are T2. 
The defense-primary \emph{Retrieval Pivot}~\cite{thornton2026retrievalpivotattacks} is included only for its secondary S4 contribution.

\subsection{Summary and Observations} \label{sec:attack_summary}
Across 64 attack-primary papers, the primary-surface distribution is S1: 42, S2: 5, S3: 1, and S4: 16. 
O1 dominates S1--S3, O2 centers on refusal or degradation, and S4 captures O3.
T2 accounts for 31 attacks, spanning generalized poisoning and aggregate extraction over unknown future queries or initially unknown records; the other 33 are T1.

\FloatBarrier
\section{Defenses and Remediation Mechanisms} \label{sec:defense}
We organize defenses by their primary control point along the knowledge-access pipeline: knowledge-base integrity and remediation (L1), retrieval-time hardening (L2), post-retrieval isolation and robust generation (L3), and access control and confidentiality (L4). Methods with the same control mechanism are grouped. Figure~\ref{fig:taxonomy_defense} gives the full taxonomy.

\begin{figure*}[tp]
\centering
\small
\tikzset{
    basic/.style  = {draw, text width=2cm, align=center, font=\sffamily, rectangle},
    root/.style   = {basic, rounded corners=2pt, thin, align=center, fill=white,text width=7.2cm, rotate=90, font=\footnotesize},
    mnode_1/.style = {basic, thin, rounded corners=2pt, align=center, fill=evalblue,text width=2.25cm, font=\footnotesize},
    mnode/.style = {basic, thin, rounded corners=2pt, align=center, fill=evalblue,text width=2.60cm, font=\footnotesize},
    wnode/.style = {basic, thin, rounded corners=2pt, align=left, fill=white,text width=9.15cm, font=\footnotesize},
}
\begin{adjustbox}{max width=\textwidth,max totalheight=0.82\textheight}
\begin{forest}
for tree={
    grow=east, reversed, growth parent anchor=east, parent anchor=east, child anchor=west,
    l sep=0.20cm, s sep=0.55mm,
    edge path={\noexpand\path[\forestoption{edge},->, >={latex}]
         (!u.parent anchor) -- +(5pt,0pt) |- (.child anchor)
         \forestoption{edge label};}
}
[RAG Defense and Remediation Mechanisms, root, parent anchor=south
    [Knowledge-Base \\ Integrity \& \\ Remediation (L1), mnode_1
        [Provenance / Forensics, mnode
            [{\renewcommand{\arraystretch}{0.94}\begin{tabular}{@{}p{4.44cm}@{\hspace{0.08cm}}p{4.44cm}@{}}
            D-RAG~\cite{e_andersen2025d}\textsuperscript{\SecTag{L2/L4},\OoneM} & Proof-Carrying Answers~\cite{shukla2025proof}\textsuperscript{\SecTag{L3},\OoneM}\\
            RAGForensics~\cite{zhang2025traceback}\textsuperscript{\OoneM} & Who Taught the Lie?~\cite{zhang2026whotaught}\textsuperscript{\OoneM}
            \end{tabular}}, wnode]
        ]
    ]
    [Retrieval-Time \\ Access \\ Hardening (L2), mnode_1
        [Reliability / Filtering, mnode
            [{\renewcommand{\arraystretch}{0.94}\begin{tabular}{@{}p{4.44cm}@{\hspace{0.08cm}}p{4.44cm}@{}}
            ReliabilityRAG~\cite{shenreliabilityrag}\textsuperscript{\SecTag{L3},\OoneM} & RA-RAG~\cite{hwang2025retrieval}\textsuperscript{\SecTag{L3},\OoneM}\\
            TrustRAG~\cite{zhou2025trustrag}\textsuperscript{\SecTag{L3},\OoneM} & GRADA~\cite{zheng2025grada}\textsuperscript{\OoneM}\\
            RAGPart/RAGMask~\cite{pathmanathan2025ragpart}\textsuperscript{\OoneM} & {RAGuard-Defense}~\cite{cheng2025secure}\textsuperscript{\OoneM}\\
            SeCon-RAG~\cite{si2025secon}\textsuperscript{\OoneM} & RAGDefender~\cite{kim2025rescuing}\textsuperscript{\OoneM}
            \end{tabular}}, wnode]
        ]
        [Detection / Secure Retrieval, mnode
            [{\renewcommand{\arraystretch}{0.94}\begin{tabular}{@{}p{4.44cm}@{\hspace{0.08cm}}p{4.44cm}@{}}
            FRAG~\cite{zhao2024frag}\textsuperscript{\SecTag{L4},\OthreeM} & GMTP~\cite{kim2025gmtp}\textsuperscript{\OoneM}\\
            EcoSafeRAG~\cite{yao2025ecosaferag}\textsuperscript{\OoneM} & Knowledge Checking~\cite{zeng2025knowledgechecking}\textsuperscript{\SecTag{L3},\OoneM}\\
            \textcolor{gray}{GGPP~\cite{hu2024ggpp}}\textsuperscript{\PrimaryA,\OoneM} & 
            \end{tabular}}, wnode]
        ]
    ]
    [Post-Retrieval \\ Context \\ Isolation (L3), mnode_1
        [Detection / Isolation, mnode
            [{\renewcommand{\arraystretch}{0.94}\begin{tabular}{@{}p{4.44cm}@{\hspace{0.08cm}}p{4.44cm}@{}}
            RobustRAG~\cite{xiang2024certifiably}\textsuperscript{\OoneM} & MMA-RAGT~\cite{mmaragt2026}\textsuperscript{\SecTag{L2/L1},\OoneM}\\
            RevPRAG~\cite{tan-etal-2025-revprag}\textsuperscript{\OoneM} & AV Filter~\cite{choudhary2026through}\textsuperscript{\OoneM}\\
            FilterRAG/ML-FilterRAG~\cite{edemacu2025defending}\textsuperscript{\OoneM} & RAGShield~\cite{patil2026ragshield}\textsuperscript{\SecTag{L1/L2},\OoneM}\\
            InstructDetector~\cite{wen-etal-2025-defending}\textsuperscript{\OoneM} & SDAG~\cite{dekel2026addressing}\textsuperscript{\SecTag{S1},\OoneM}
            \end{tabular}}, wnode]
        ]
        [Robust Generation, mnode
            [{\renewcommand{\arraystretch}{0.94}\begin{tabular}{@{}p{4.44cm}@{\hspace{0.08cm}}p{4.44cm}@{}}
            ATM~\cite{zhu2024atm}\textsuperscript{\OoneM} & INFO-RAG~\cite{xu2024inforag}\textsuperscript{\OoneM}\\
            Discern-and-Answer~\cite{hong2024so}\textsuperscript{\OoneM} & InstructRAG~\cite{weiinstructrag}\textsuperscript{\OoneM}\\
            Astute RAG~\cite{wang2025astute}\textsuperscript{\OoneM} & RbFT~\cite{tu2025rbft}\textsuperscript{\OoneM}\\
            Credibility-Aware Generation~\cite{pan2024credibility}\textsuperscript{\SecTag{L1},\OoneM} & Fact-Centric Alignment~\cite{hao2025factcentric}\textsuperscript{\OoneM}\\
            RPO~\cite{yan2025rpo}\textsuperscript{\OoneM} & \textcolor{gray}{ABDD~\cite{li2025llms}}\textsuperscript{\PrimaryE,\OoneM}
            \end{tabular}}, wnode]
        ]
        [Disclosure Controls, mnode
            [{\renewcommand{\arraystretch}{0.94}\begin{tabular}{@{}p{4.44cm}@{\hspace{0.08cm}}p{4.44cm}@{}}
            SD-RAG~\cite{masoud2026sd}\textsuperscript{\SecTag{L2},\OthreeM} & PAD~\cite{wang2025privacy}\textsuperscript{\OthreeM}\\
            PPRAG~\cite{zhang2026privacy}\textsuperscript{\SecTag{L2},\OthreeM} & SEAL-Tag~\cite{xie2026seal}\textsuperscript{\SecTag{L4},\OthreeM,\TtwoP}
            \end{tabular}}, wnode]
        ]
    ]
    [{Access Control, \\ Privacy \& \\ Confidentiality (L4)}, mnode_1
        [Authorization / Membership, mnode
            [{\renewcommand{\arraystretch}{0.94}\begin{tabular}{@{}p{4.44cm}@{\hspace{0.08cm}}p{4.44cm}@{}}
            AC-RAG~\cite{10.1145/3672608.3707848}\textsuperscript{\OthreeM} & Retrieval Pivot~\cite{thornton2026retrievalpivotattacks}\textsuperscript{\SecTag{S4},\OthreeM}\\
            Mirabel~\cite{choi2025mirabel}\textsuperscript{\SecTag{L2},\OthreeM} & Ensemble Privacy Defense~\cite{fu2026ensembleprivacy}\textsuperscript{\SecTag{L3/S4},\OthreeM}
            \end{tabular}}, wnode]
        ]
        [Differential Privacy /\\Corpus Transformation, mnode
            [{\renewcommand{\arraystretch}{0.94}\begin{tabular}{@{}p{4.44cm}@{\hspace{0.08cm}}p{4.44cm}@{}}
            DPVoteRAG~\cite{koga2024privacy}\textsuperscript{\OthreeM} & RAG with Differential Privacy~\cite{grislain2025rag}\textsuperscript{\OthreeM}\\
            LPRAG~\cite{he2025mitigating}\textsuperscript{\OthreeM} & VAGUE-Gate~\cite{hemmat2025vague}\textsuperscript{\OthreeM}\\
            InvisibleInk~\cite{vinod2025invisibleink}\textsuperscript{\OthreeM} & SAGE~\cite{zeng2025mitigating}\textsuperscript{\OthreeM}\\
            DP-SynRAG~\cite{mori2026dpsynrag}\textsuperscript{\SecTag{L1},\OthreeM,\TtwoP} & AURA~\cite{wang2026making}\textsuperscript{\SecTag{L1},\OthreeM}
            \end{tabular}}, wnode]
        ]
        [Secure Backends /\\Architectures, mnode
            [{\renewcommand{\arraystretch}{0.94}\begin{tabular}{@{}p{4.44cm}@{\hspace{0.08cm}}p{4.44cm}@{}}
            RemoteRAG~\cite{cheng2025remoterag}\textsuperscript{\OthreeM} & ppRAG~\cite{ye2025efficient}\textsuperscript{\SecTag{L2},\OthreeM}\\
            FedE4RAG~\cite{mao2025privacy}\textsuperscript{\OthreeM} & C-FedRAG~\cite{addison2024c}\textsuperscript{\OthreeM}\\
            Privacy-Aware RAG~\cite{zhou2025privacy}\textsuperscript{\OthreeM} & SAG~\cite{zhou2025provably}\textsuperscript{\SecTag{L1},\OthreeM}\\
            SecureRAG~\cite{bassit2025securerag}\textsuperscript{\SecTag{L2},\OthreeM} & 
            \end{tabular}}, wnode]
        ]
    ]
]
\end{forest}
\end{adjustbox}
\caption{
    Taxonomy of RAG defense and remediation mechanisms, organized by primary control point (L1--L4).
    Superscripts mark the security objective each method counters: \textcolor{red}{O1} integrity, \textcolor{blue}{O2} availability, and \textcolor{green!60!black}{O3} confidentiality.
    \textcolor{gray}{Gray sec:} tags record explicit secondary layers or attack surfaces and are excluded from primary counts. The gray GGPP and ABDD cross-lists are attack- and evaluation-primary, respectively.
    T2 marks 2 of 56 defense-primary works with aggregate protection over unknown future queries or adaptively explored records. Retrieval Pivot is defense-primary, with S4 as a secondary audit role.
}
\label{fig:taxonomy_defense}
\Description{A hierarchical taxonomy of 56 defense-primary studies by control layer and mechanism, with objective, target-scope, and secondary-role tags.}
\end{figure*}

\subsection{Knowledge-Base Integrity, Provenance, and Remediation}\label{ssec:defense:l1}
L1 either protects the substrate before ingestion or attributes failures to stored evidence. 
For admission and provenance, \emph{D-RAG}~\cite{e_andersen2025d} decentralizes provenance and access control, 
while \emph{Proof-Carrying Answers}~\cite{shukla2025proof} binds cryptographic provenance to the evidence-to-answer path. 
For remediation, \emph{RAGForensics}~\cite{zhang2025traceback} traces malicious outputs to responsible passages, 
and \emph{Who Taught the Lie?}~\cite{zhang2026whotaught} scores responsibility from retrieval and response signals. 
With only these four L1-primary works, upstream protection remains much thinner than S1 attack coverage.

\subsection{Retrieval-Time Access Hardening}\label{ssec:defense:l2}
L2 secures retrieval before external content becomes model-visible context. 
Reliability methods reduce dependence on any single passage. For example, \emph{ReliabilityRAG}~\cite{shenreliabilityrag} aggregates a consistent majority, and \emph{RA-RAG}~\cite{hwang2025retrieval} weights source credibility.
In addition, \emph{TrustRAG}~\cite{zhou2025trustrag} and \emph{GRADA}~\cite{zheng2025grada} identify weak support through clustering or graph connectivity, 
and \emph{RAGPart/RAGMask}~\cite{pathmanathan2025ragpart} test sensitivity through partitioning and masking.

Candidate-cleaning methods try to separate poison before generation: 
\emph{RAGuard-Defense}~\cite{cheng2025secure}\footnote{Two unrelated works are both named RAGuard; we use RAGuard-Defense and RAGuard-Bench for disambiguation.} filters expanded results using perplexity and similarity,
while \emph{SeCon-RAG}~\cite{si2025secon} and \emph{RAGDefender}~\cite{kim2025rescuing} combine semantic, cluster, conflict, or purification signals. 
Detection methods use different evidence: 
\emph{GMTP}~\cite{kim2025gmtp} uses masked-token probabilities, 
while \emph{EcoSafeRAG}~\cite{yao2025ecosaferag} and \emph{Knowledge Checking}~\cite{zeng2025knowledgechecking} compare contextual or representation-level signals. 
By contrast, \emph{FRAG}~\cite{zhao2024frag} differs in objective, protecting federated vector retrieval against O3 leakage. 
The attack-primary \emph{GGPP}~\cite{hu2024ggpp} is cross-listed for its secondary detector.

L2 is substantial, but aggregation often assumes a bounded malicious minority and filtering assumes that poison is anomalous, weakly supported, or separable. 
Fluent, retrievable, corpus-fitting payloads increasingly weaken these assumptions.

\subsection{Post-Retrieval Isolation and Robust Generation}\label{ssec:defense:l3}
L3 acts after harmful content enters the retrieved context. 
Detection methods inspect the passages or model behavior: 
\emph{RevPRAG}~\cite{tan-etal-2025-revprag} uses activations, \emph{AV Filter}~\cite{choudhary2026through} uses attention variance, and \emph{FilterRAG/ML-FilterRAG}~\cite{edemacu2025defending} learn corpus-level or statistical filters. 
\emph{RAGShield}~\cite{patil2026ragshield} checks numerical claims across sources and time, while \emph{InstructDetector}~\cite{wen-etal-2025-defending} identifies embedded instructions. 
\emph{RobustRAG}~\cite{xiang2024certifiably} instead isolates passages and certifies the answer under bounded corruption.

Another way is to constrain interaction among retrieved documents. 
Isolation methods include \emph{SDAG}~\cite{dekel2026addressing}, which uses sparse document attention, and \emph{MMA-RAGT}~\cite{mmaragt2026}, which extends stateful trust isolation to multimodal agentic RAG.
Robust-generation methods discount misleading evidence in different ways: \emph{Discern-and-Answer}~\cite{hong2024so} removes it, 
\emph{InstructRAG}~\cite{weiinstructrag} synthesizes denoising rationales, 
\emph{Astute RAG}~\cite{wang2025astute} reconciles internal and retrieved knowledge, 
and \emph{RbFT}~\cite{tu2025rbft} fine-tunes against misleading or counterfactual context. 
In addition, \emph{ATM}~\cite{zhu2024atm} and \emph{INFO-RAG}~\cite{xu2024inforag} train adversarial or unsupervised refinement, 
while \emph{Credibility-Aware Generation}~\cite{pan2024credibility}, \emph{Fact-Centric Alignment}~\cite{hao2025factcentric}, and \emph{RPO}~\cite{yan2025rpo} align generation at the source, fact, or preference level. 
The evaluation-primary \emph{ABDD} study (\emph{LLMs Trust Humans More})~\cite{li2025llms} is cross-listed for its CDEQ mitigation as a secondary defense contribution.

A parallel L3 line controls disclosure after retrieval. 
\emph{SD-RAG}~\cite{masoud2026sd} sanitizes context, \emph{PAD}~\cite{wang2025privacy} protects decoding, and \emph{PPRAG}~\cite{zhang2026privacy} combines privacy-aware retrieval with context transformation, 
while \emph{SEAL-Tag}~\cite{xie2026seal} aggregates evidence tags across query sequences to limit linkability and mosaic disclosure, giving T2 coverage. 
L3 is broad, but many methods require model access, extra inference, or detectable abnormal signals.

\subsection{Access Control, Privacy, and Confidentiality}\label{ssec:defense:l4}
L4 controls who can access retrieved knowledge and what can be inferred from it. 
Authorization and membership defenses include \emph{AC-RAG}~\cite{10.1145/3672608.3707848}, which integrates fine-grained authorization, and \emph{Retrieval Pivot}~\cite{thornton2026retrievalpivotattacks}, which enforces authorization at each graph-traversal hop; Retrieval Pivot is defense-primary, with its S4 leakage audit as a secondary role.
In addition, \emph{Mirabel}~\cite{choi2025mirabel} and \emph{Ensemble Privacy Defense}~\cite{fu2026ensembleprivacy} suppress membership signals used to infer corpus membership.

Privacy mechanisms limit disclosure from text, retrieval, or generation. 
\emph{DPVoteRAG}~\cite{koga2024privacy}, \emph{RAG with Differential Privacy}~\cite{grislain2025rag}, \emph{LPRAG}~\cite{he2025mitigating}, \emph{VAGUE-Gate}~\cite{hemmat2025vague}, and \emph{InvisibleInk}~\cite{vinod2025invisibleink} apply differential privacy at token, entity, retrieval, or generation level. 
\emph{SAGE}~\cite{zeng2025mitigating} replaces private corpora with synthetic alternatives, 
\emph{DP-SynRAG}~\cite{mori2026dpsynrag} creates a reusable differentially private corpus under a one-time budget (T2), 
and \emph{AURA}~\cite{wang2026making} adulterates proprietary graphs so extracted copies lose utility.

Backend methods protect retrieval through architecture or cryptography. 
For instance, \emph{RemoteRAG}~\cite{cheng2025remoterag} secures cloud retrieval and \emph{ppRAG}~\cite{ye2025efficient} uses distance-preserving encryption. 
Additionally, \emph{FedE4RAG}~\cite{mao2025privacy}, \emph{C-FedRAG}~\cite{addison2024c}, and \emph{Privacy-Aware RAG}~\cite{zhou2025privacy} use federated or isolated retrieval, 
while \emph{SAG}~\cite{zhou2025provably} adds formal security and \emph{SecureRAG}~\cite{bassit2025securerag} provides an end-to-end protected architecture. 
These controls commonly trade utility, latency, or retrieval quality for protection.

\subsection{Summary and Observations}\label{ssec:defense:summary}
Among 56 defense-primary papers, L1: 4, L2: 12, L3: 21, and L4: 19. 
The stage mismatch is largest upstream: 42 attacks begin at S1, but only four defenses begin at L1. 
The target mismatch is also sharp: only DP-SynRAG and SEAL-Tag are T2 (2/56), while the other 54 defenses are T1. 
Finally, the signal mismatch persists because fluent, retrievable S1 payloads weaken the anomaly and separability assumptions common in L2/L3. 
Section~\ref{sec:insights} will further revisit these three mismatches.

\FloatBarrier
\section{Secure-RAG Evaluation Studies} \label{sec:eval}

\begin{table}[!t]
    \caption{
        Coverage map of secure-RAG evaluation studies.
        \textcolor{gray}{Gray cross-lists} show their primary class (A: attack; D: defense); unmarked works are evaluation-primary.
        \textbf{Objective}: O1 integrity, O2 availability, O3 confidentiality;
        \textbf{Target}: T1 fixed/known, T2 aggregate/cross-query;
        \textbf{Surface}: S1 corpus/indexing, S2 retrieval, S3 context/prompt, S4 extraction.
        \protect\cmark/$\circ$/-- denote central, partial, and not primary coverage, respectively.
    }
    \label{tab:bench_evaluation}
    \centering
    \small
    \setlength{\tabcolsep}{2.2pt}
    \renewcommand{\arraystretch}{0.9}
    \begin{tabular}{>{\raggedright\arraybackslash}p{2.8cm} ccc cc cccc}
        \toprule
        \multirow{2}{*}{\textbf{Paper}}
            & \multicolumn{3}{c}{\textbf{Objective}}
            & \multicolumn{2}{c}{\textbf{Target}}
            & \multicolumn{4}{c}{\textbf{Surface}} \\
        \cmidrule(lr){2-4} \cmidrule(lr){5-6} \cmidrule(lr){7-10}
            & O1 & O2 & O3 & T1 & T2 & S1 & S2 & S3 & S4 \\
        \midrule
        \multicolumn{10}{l}{\textit{Benchmark Studies}} \\
        Rag~'n~Roll \citeyearpar{de2024rag}           & \cmark  & --      & --      & \cmark  & -- & \cmark  & \cmark  & \cmark  & --      \\
        BIPIA \citeyearpar{yi2025benchmarking}       & \cmark  & $\circ$      & --      & \cmark  & -- & --      & --      & \cmark  & --      \\
        Generative-Search Robustness \citeyearpar{hu2024evaluating} & \cmark & -- & -- & \cmark & -- & -- & -- & -- & -- \\
        ReEval \citeyearpar{yu2024reeval}             & \cmark  & --      & --      & \cmark  & -- & --      & --      & \cmark  & --      \\
        SafeRAG \citeyearpar{liang2025saferag}        & \cmark  & \cmark & --      & \cmark  & -- & \cmark  & \cmark  & \cmark  & --      \\
        RSB \citeyearpar{zhang2025benchmarking}      & \cmark  & \cmark & --      & \cmark  & -- & \cmark  & $\circ$ & $\circ$ & --      \\
        PoisonArena \citeyearpar{chen2025poisonarena} & \cmark & --     & --      & \cmark  & $\circ$ & \cmark  & \cmark  & --      & --      \\
        {RAGuard-Bench} \citeyearpar{zeng2026worse}           & \cmark  & --      & --      & \cmark  & -- & $\circ$ & --      & --      & --      \\
        OpenRAG-Soc \citeyearpar{guo2026hidden}      & \cmark  & --      & --      & \cmark  & -- & \cmark  & $\circ$ & \cmark  & --      \\
        MPIB \citeyearpar{lee2026mpib}               & \cmark  & --      & --      & \cmark  & -- & --      & --      & \cmark  & --      \\
        \textcolor{gray}{S-RAG \citeyearpar{zeng2025s}}\textsuperscript{\PrimaryA}                 & --      & --      & \cmark  & \cmark  & -- & --      & --      & --      & \cmark  \\
        \textcolor{gray}{SMA \citeyearpar{sun2025sma}}\textsuperscript{\PrimaryA}                  & --      & --      & \cmark  & \cmark  & -- & --      & --      & --      & \cmark  \\
        \textcolor{gray}{Privacy Protection in RAG \citeyearpar{zhang2026privacy}}\textsuperscript{\PrimaryD} & -- & -- & \cmark & \cmark & -- & -- & -- & -- & \cmark \\
        KE-Bench \citeyearpar{qi2026benchmarking}    & --      & --      & \cmark  & {--} & {\cmark} & --      & $\circ$ & --      & \cmark  \\
        MedPriv-Bench \citeyearpar{guan2026medpriv}  & --      & --      & \cmark  & \cmark  & -- & --      & --      & --      & \cmark  \\
        \textcolor{gray}{SEAL-Tag \citeyearpar{xie2026seal}}\textsuperscript{\PrimaryD}            & $\circ$ & -- & \cmark & $\circ$ & \cmark & -- & -- & -- & \cmark \\
        \midrule
        \multicolumn{10}{l}{\textit{Systematic Evaluation Studies}} \\
        Instruction-Following Retrievers \citeyearpar{behnamghader2025exploiting} & \cmark & -- & -- & \cmark & -- & -- & \cmark & -- & -- \\
        RAG LLMs are Not Safer \citeyearpar{an2025rag}                       & \cmark & -- & -- & \cmark & -- & -- & $\circ$ & -- & -- \\
        ABDD \citeyearpar{li2025llms}                           & \cmark & -- & -- & \cmark & -- & -- & -- & -- & -- \\
        \textcolor{gray}{Good\&Bad \citeyearpar{zeng2024good}}\textsuperscript{\PrimaryA}         & --      & --      & \cmark  & \cmark  & -- & --      & $\circ$ & $\circ$ & \cmark  \\
        \textcolor{gray}{Beyond Text \citeyearpar{zhang2025beyond}}\textsuperscript{\PrimaryA}     & --      & --      & \cmark  & \cmark  & -- & --      & $\circ$ & $\circ$ & \cmark  \\
        \textcolor{gray}{mRAG Privacy \citeyearpar{al2026systemic}}\textsuperscript{\PrimaryA}              & --      & --      & \cmark  & \cmark  & -- & --      & --      & --      & \cmark  \\
        \bottomrule
    \end{tabular}

\end{table}

Within the scope of Section~\ref{ssec:scope}, \emph{benchmark studies} provide reusable resources, whereas \emph{systematic evaluation studies} characterize secure-RAG failures.
Table~\ref{tab:bench_evaluation} maps 15 evaluation-primary works and attack/defense cross-lists (excluded from the count) to SLOT.

\subsection{Benchmark Studies}

\noindent\textbf{Integrity (O1).}
The largest cluster tests whether adversarial or misleading inputs compromise answer integrity.
\emph{Rag 'n Roll}~\cite{de2024rag} evaluates indirect prompt manipulation in end-to-end applications rather than isolated components.
\emph{Generative-Search Robustness}~\cite{hu2024evaluating} black-box tests generative search systems with adversarial factoid questions, while \emph{ReEval}~\cite{yu2024reeval} replaces answers in coherent evidence and adds relevant distractors for transferable stress tests.
\emph{PoisonArena}~\cite{chen2025poisonarena} models multi-adversary competition, compares defenses, and provides partial T2 coverage through transfer to held-out equivalent queries.
{\emph{RAGuard-Bench}}~\cite{zeng2026worse} uses naturally occurring web misinformation rather than synthetic poison; \emph{OpenRAG-Soc}~\cite{guo2026hidden} jointly tests injection, poisoning, and defenses on web/social corpora; and \emph{MPIB}~\cite{lee2026mpib} compares indirect-injection harms and defenses across a range of application tasks.

\noindent\textbf{Integrity and availability jointly (O1, O2).}
Availability rarely appears as a first-class target; it is mostly covered as a single task within broader poisoning suites.
\emph{BIPIA}~\cite{yi2025benchmarking} benchmarks indirect prompt injection across application tasks, attack goals, and injection positions, and evaluates boundary-aware and explicit-reminder defenses in black- and white-box settings.
\emph{SafeRAG}~\cite{liang2025saferag} injects Noise, Conflict, Toxicity, and DoS at each pipeline stage and exposes 14 component-level evaluations.
\emph{RAG Security Bench (RSB)}~\cite{zhang2025benchmarking} evaluates 13 poisoning types, spanning targeted, DoS, and trigger-DoS attacks, against defenses across sequential, branching, conditional, looping, multi-turn, multimodal, and agentic RAG.

\noindent\textbf{Confidentiality (O3).}
A growing line evaluates leakage from the retrieval corpus, a property specific to RAG and distinct from generic prompt leakage.
\emph{S-RAG}~\cite{zeng2025s} audits, under black-box access, whether a user's personal data has been ingested into the index.
\emph{SMA}~\cite{sun2025sma} introduces source-aware membership auditing to semi-black-box and multimodal retrieval.
\emph{Privacy Protection in RAG}~\cite{zhang2026privacy} introduces a knowledge-graph-based, fine-grained privacy processor and evaluates extraction and personal identification alongside answer utility.
\emph{KE-Bench}~\cite{qi2026benchmarking} standardizes knowledge-extraction evaluation across retrievers, generators, and datasets, with RandText, DGEA, CopyBreak, and IKEA baselines for aggregate recovery of initially unknown datastore content (T2).
\emph{MedPriv-Bench}~\cite{guan2026medpriv} jointly evaluates medical contextual leakage and clinical utility across three privacy-preserving methods.
\emph{SEAL-Tag}~\cite{xie2026seal} combines a PII-Evidence Table with probabilistic-circuit routing and evaluates direct, linkability, adaptive-extraction, and PET-Spoofing attacks alongside utility and latency.

\subsection{Systematic Evaluation Studies}
Beyond benchmark construction, systematic studies clarify what secure-RAG should measure.
On integrity, \emph{Instruction-Following Retrievers}~\cite{behnamghader2025exploiting} measures how malicious instructions exploit instruction-following retrievers at S2 to surface attacker-preferred information, while \emph{RAG LLMs are Not Safer}~\cite{an2025rag} compares RAG and non-RAG safety, showing that retrieved context can induce unsafe outputs even when the model and documents are individually safe, and evaluates retriever-aware red teaming.
\emph{ABDD}~\cite{li2025llms} introduces the Authority Bias Detection Dataset to measure bias under conflicting user-provided and retrieved evidence, and proposes the Conflict Detection Enhanced Query (CDEQ) mitigation.
It is evaluation-primary because measurement is central, with CDEQ as a secondary defense contribution.

On confidentiality, \emph{Good\&Bad}~\cite{zeng2024good} shows that retrieval introduces a corpus-leakage channel but can mitigate certain parametric-memorization risks.
\emph{Beyond Text}~\cite{zhang2025beyond} demonstrates private-information extraction across vision- and speech-language RAG.
{\emph{mRAG Privacy}}~\cite{al2026systemic} evaluates image membership inference and image-caption retrieval attacks in vision-centric mRAG.
Together, they show where leakage appears and how it changes across modalities and assumptions.

\subsection{Limitations of Current Evaluation} \label{sec:eval_limits}
Coverage is uneven along the axes of our threat model, and the gaps below cluster around the insights we revisit in Section~\ref{sec:insights}.

\noindent\textbf{Target axis: T1 dominance.}
{Of the 15 evaluation-primary works, KE-Bench is the only study centered on T2; PoisonArena provides partial T2 coverage through held-out query variants. The other 14 use fixed instances or query sets as their primary setting.}
{T2 evaluation must aggregate stance shift, refusal or degradation, or recovered content over a query distribution or interaction sequence.}

\noindent\textbf{Surface axis: multi-surface coverage without a unified harness.}
\emph{SafeRAG}~\cite{liang2025saferag}, \emph{OpenRAG-Soc}~\cite{guo2026hidden}, and \emph{Rag 'n Roll}~\cite{de2024rag} already exercise multiple attack surfaces or pipeline stages.
There is no unified, defense-aware, pluggable cross-surface harness that exposes common interfaces for {S1--S4 and L1--L4}, leaving cross-surface interactions and failure propagation difficult to evaluate.

\noindent\textbf{Adaptive evaluation {is} rarely reported.}
Few defenses (notably RobustRAG~\cite{xiang2024certifiably} and RAGShield~\cite{patil2026ragshield}) are evaluated against attackers aware of the defense.
Across studies, headline attack success rate (ASR), distribution-level stance shift, and certified accuracy are not directly rankable because they evaluate different targets under different attacker budgets and assumptions.
Section~\ref{ssec:evaluation_protocol} details these conditions.

\noindent\textbf{No standardized defense harness.}
The two {defense-primary works cross-listed here} (Privacy Protection in RAG~\cite{zhang2026privacy}, SEAL-Tag~\cite{xie2026seal}) center on the defenses they propose, while benchmarks that compare multiple defenses (BIPIA~\cite{yi2025benchmarking}, RSB~\cite{zhang2025benchmarking}, PoisonArena~\cite{chen2025poisonarena}, OpenRAG-Soc~\cite{guo2026hidden}, MPIB~\cite{lee2026mpib}, KE-Bench~\cite{qi2026benchmarking}, and MedPriv-Bench~\cite{guan2026medpriv}) evaluate them under benchmark-specific threat protocols.
No shared harness exists for comparing arbitrary defenses against a standardized attack suite, leaving each defense paper to set its own conditions and making cross-defense comparison unreliable.
{Our data release contains the screening process, selected-paper metadata, and machine-readable SLOT labels; Section~\ref{ssec:evaluation_protocol} provides a protocol specification, not executable evaluation software.}

\noindent\textbf{Confidentiality metrics {are} incommensurable across studies and modalities.}
The O3 cluster now spans reusable privacy audits, leakage benchmarks, and attack-defense evaluations, but each study reports leakage in incompatible units (membership AUC, exact-match extraction, PII recall, etc.), preventing cross-study severity comparison.
The gap widens across modalities, where no shared interface-level leakage rate exists.

\subsection{Evaluation Protocol and Reporting Specification}

\label{ssec:evaluation_protocol}

{Our proposed six-stage plug-in harness exposes} the external-source, ingestion/indexing, retrieval/reranking, context-assembly, generation, and response/remediation stages. Attacks plug into S1--S4 and defenses into L1--L4; each result is reported in the relevant O$\times$T cell. A surface-complete stack must contain an explicit control for every S1--S4 surface and evaluate each control separately against an attacker who knows all non-secret defense details.

The minimum SLOT card reports S/L/O/T, attacker knowledge and access, a surface-specific budget, victim configuration, assumptions, adaptive status, utility, latency, and uncertainty. O1 evaluation should report task success and integrity under attack; O2 should report refusal/failure and resource costs; O3 should report membership or leakage together with recovered fraction and query cost. T1 fixes the target instances or query set before the attack; T2 measures aggregate behavior over unknown future queries or adaptively explored records in the underlying knowledge corpus.

\begin{table*}[t]
    \caption{Comparison of representative secure-RAG evaluations under their primary SLOT labels. Secondary labels are omitted for clarity; for BIPIA, S3 and L3 denote its cross-tagged attack and defense roles, respectively.}
    \label{tab:evaluation_examples}
    \centering
    \footnotesize
    \setlength{\tabcolsep}{5pt}
    \renewcommand{\arraystretch}{1.4}
    \begin{tabularx}{\textwidth}{@{}>{\raggedright\arraybackslash}p{0.20\textwidth} >{\raggedright\arraybackslash}p{0.35\textwidth} >{\raggedright\arraybackslash}X@{}}
        \toprule
        \textbf{Work}
        & \textbf{Threat model, access, and budget}
        & \textbf{Representative evaluation} \\
        \midrule

        PoisonedRAG~\cite{zou2025poisonedrag}
        (S1/O1/T1)
        & Target query and answer known; corpus-write access; five injected passages per query.
        & NQ, HotpotQA, and MS MARCO; 97\% attack success rate (ASR) on NQ with five injected passages per query. \\

        Topic-FlipRAG~\cite{gong2025topicfliprag}
        (S1/O1/T2)
        & Black-box topic/stance manipulation; exact future test queries need not be known.
        & PROCON and MS MARCO; distribution-level attack success value (ASV) of 0.49 for Qwen2.5 and 0.48 for Llama3.1 on PROCON. \\

        RobustRAG~\cite{xiang2024certifiably}
        {(L3/O1/T1)}
        & Defense-aware bounded corruption; certification requires benign evidence to dominate malicious evidence.
        & RQA-MC, RQA, NQ, and Biography; 69--71\% certifiable robust accuracy on RQA-MC and 31--38\% on RQA under bounded corruption. \\

        BIPIA~\cite{yi2025benchmarking}
        (S3/L3/O1/T1)
        & Indirect prompt injection through external content; attack success and clean utility are measured separately.
        & Five tasks, 250 attacker goals, and 25 LLMs; 626,250 training and 86,250 test prompts, evaluated using ASR and utility metrics. \\

        \bottomrule
    \end{tabularx}

\end{table*}

Table~\ref{tab:evaluation_examples} illustrates why cross-paper results require protocol-level alignment by comparing four representative studies spanning targeted corpus poisoning, distribution-level stance manipulation, certified defense, and indirect prompt injection.

As these studies evaluate different targets under different attacker access, budgets, assumptions, and metrics, their headline ASR, distribution-level ASV, and certifiable robust accuracy are not directly rankable.

\section{Insights and Future Directions}\label{sec:insights}
The SLOT map exposes three mismatches and motivates six directions: close the target, stage, and signal gaps (I1--I3); build surface-complete defenses (I4); strengthen confidentiality controls (I5); and evaluate multimodal and agentic RAG (I6).

\paragraph{I1: Study more realistic targets.}
T2 accounts for 31/64 attack-primary works, but only 2/56 defense-primary works and 1/15 evaluation-primary studies. 
Therefore, defenses and evaluations should audit reliability across a query distribution or interaction sequence (T2), aggregating stance shift (O1), refusal or degradation (O2), or recovered content and query cost (O3).

\paragraph{I2: Close the stage mismatch.}
Although 42 attacks begin at S1, only 4 defenses begin at L1. 
Thus, corpus admission, provenance, monitoring, and remediation require more attention.

\paragraph{I3: Close the signal mismatch.}
Recent S1 attacks produce fluent, retrievable, corpus-fitting payloads, while many L2/L3 defenses still look for semantic anomalies.
Defenses should therefore build on provenance, cross-query behavior, or causal influence rather than surface abnormality.

\paragraph{I4: Build surface-complete defenses.}
A surface-complete stack places an explicit control against every S1--S4 surface and tests each control against an adaptive attacker who knows all non-secret defense details. A shared harness should fix surface-specific budgets and report a compact SLOT card covering S/L/O/T, attacker access, victim setup, utility, latency, and uncertainty; Section~\ref{ssec:evaluation_protocol} provides the template.

\paragraph{I5: Strengthen confidentiality controls.}
Most confidentiality defenses remain T1 and rely on differential privacy, encryption, or architectural change. T2 evaluations should additionally report recovered corpus fraction and query cost alongside policy violations, utility, and latency.

\paragraph{I6: Evaluate multimodal and agentic RAG.}
Multimodal leakage and complex poisoning already span vision, speech, multi-turn, and agentic RAG~\cite{zhang2025beyond,al2026systemic,zhang2025benchmarking}. \emph{AgentPoison}, \emph{MemoryGraft}, and \emph{MINJA} further show that poisoned memory can affect later queries or actions across sessions~\cite{chen2024agentpoisonredteamingllmagents,srivastava2025memorygraftpersistentcompromisellm,dong2026memoryinjectionattacksllm}. Evaluation should therefore test persistence across modalities, turns, and sessions, including whether a payload triggers an action in a later interaction.

\section{Conclusion}\label{sec:conclusion}

We surveyed secure RAG within an external-knowledge boundary that separates RAG-introduced or RAG-amplified risks from inherent LLM flaws. 
We conducted a systematic literature search and curated 135 works. We release the screening process, selected-paper metadata, and machine-readable SLOT labels.
The synthesis clarifies which failures fall within secure RAG and reveals three mismatches: a target mismatch (31/64 T2 attacks versus 2/56 T2 defenses and 1/15 T2-centered evaluation studies), a stage mismatch (42 primary-S1 attacks versus 4 primary-L1 defenses), and a signal mismatch (fluent, retrievable, corpus-fitting S1 attacks versus L2/L3 defenses that often assume anomalous or separable poison). 
We discussed surface-complete defense stacks, comparable evaluation, practical confidentiality controls, and persistence-aware multimodal and agentic testing as directions for closing these gaps.

\balance
\section*{Limitations}
First, secure RAG is a fast-moving area, and our review is a snapshot through April 30, 2026. Later work may change the observed distribution. 
Second, LLM-assisted screening and expert selection can introduce errors, while English-language and full-text availability constraints may underrepresent work from some communities. 
Finally, our tables and observations provide a qualitative synthesis rather than a precise meta-analysis, and should be read as highlighting structural patterns rather than exact field-level measurements.

\section*{Ethical Considerations}
This work is a literature survey and introduces no new attacks, defenses, or experiments. All included material is previously published and cited to the original sources.
This work introduces no new exploit code or undisclosed vulnerabilities. 
Although machine assistance supported screening, experts made the final inclusion and SLOT-tagged decisions, and venue prestige was not a selection criterion. 
We recognize the dual-use risk of organizing attack methods in one place; we mitigate it by emphasizing threat assumptions, crossed boundaries, and corresponding controls. 
The released screening process, selected-paper metadata, and machine-readable SLOT labels are intended to support reproducible analysis and stronger protection research for RAG systems.

\section*{Acknowledgments}
This work was partially supported by the National Key R\&D Program of China under Grant No. 2023YFF0725100; NSFC under Grant No. U22B2060; Guangdong Basic and Applied Basic Research Foundation under Grant No. 2026A1515010227; PolyU TDG under Grant No. TDG25-28/TBP/11-IICA; PolyU Internal Research Fund (No. BDZ3); PolyU LTC Project (Grant TDLEG25-28/IICA/P/05, No. 48EM); NSFC/RGC Joint Research Scheme under Grant No. N\_PolyU5179/25; Hong Kong RGC under Grant No. PolyU25600624; ITF under Grant Nos. ITS/052/23MX and PRP/009/22FX; industrial sponsors; and RMGS under Grant Nos. P0045948, P0048183, P0048191, P0046453, and P0060272.

\bibliography{survey}
\clearpage
\nobalance
\appendix
\pagestyle{empty}
\thispagestyle{empty}

\section{Search and Screening Protocol}
\label{app:search_protocol}

\paragraph{Sources, window, and queries.}
We searched arXiv, ACL Anthology, OpenReview, Semantic Scholar, and DBLP for work first made public from May 1, 2020 through April 30, 2026, inclusive. 
Each source query combined at least one RAG term with at least one security term in the title or abstract. 

The RAG block was as follows:
\begin{quote}\small\raggedright\ttfamily
retrieval augmented OR retrieval-augmented OR RAG OR GraphRAG OR
agentic RAG OR multimodal RAG OR
graph retrieval augmented generation OR
multimodal retrieval augmented generation OR
visual document retrieval augmented generation
\end{quote}

The security block was as follows:
\begin{quote}\small\raggedright\ttfamily
attack* OR adversarial OR poison* OR backdoor* OR "prompt injection" OR
jailbreak* OR hijack* OR malicious OR threat* OR vulnerab* OR compromis* OR
manipulat* OR corrupt* OR perturb* OR exploit* OR privacy OR private OR
confidential* OR leak* OR "data extraction" OR "knowledge extraction" OR
"membership inference" OR unauthoriz* OR "access control" OR provenance OR
copyright OR watermark* OR steal* OR theft OR protect* OR security OR secure OR
defen* OR robust* OR safe* OR trust* OR reliab* OR integrity OR verif* OR
audit* OR guardrail* OR resilien*
\end{quote}

Because the DBLP publication-search API does not expose abstracts, both blocks were matched against DBLP titles only.

\paragraph{Title-and-abstract prompt.}
The first screening round used the following instruction with the returned title and abstract:
\begin{quote}
Decide whether the external-knowledge path of a generative system is the object of a primary or substantial contribution. Include (I1) attacks on a corpus, index, retriever, ranking, memory, graph, tool content, or retrieved evidence, including indirect prompt injection and extraction; (I2) defenses, secure architectures, access control, selective disclosure, cryptographic protection, or remediation against these risks; (I3) security/privacy benchmarks, red-team studies, surveys, or measurements of behavior under adversarial or untrustworthy retrieved evidence; (I4) methods or evaluations centered on the integrity of false, misleading, conflicting, contaminated, or temporally invalid external evidence; and (I5) query or prompt perturbations designed to manipulate retrieval or a retrieval-backed generator. Exclude ordinary RAG accuracy, retrieval optimization, efficiency, citation quality, benign irrelevant context, generic hallucination reduction, generic LLM safety/privacy, unrelated domain applications, and broad risk discussions without a concrete external-knowledge contribution. Return INCLUDE, EXCLUDE, or UNCERTAIN, the matched criterion, surface, and a brief evidence-based reason.
\end{quote}

\paragraph{Full-text prompt.}
The second round applied the following rule to full-text evidence:
\begin{quote}
Include only if all three conditions hold: (A) the system uses generation grounded in an external corpus, retrieved documents, graph, multimodal or code store, external memory, tool-returned content, or private inference-time context; (B) security, privacy, or semantic/source integrity of that path is a primary or substantial contribution; and (C) the paper provides a concrete attack, defense, benchmark, empirical measurement, integrity-robustness method, RAG-targeted input attack, or systematic survey/taxonomy. Apply I1--I5 as defined above. Exclude ordinary quality, efficiency, abstention, benign noise, generic LLM safety/privacy, unrelated applications, and non-concrete discussions. Use UNCERTAIN only when the extracted evidence is inadequate or genuinely ambiguous, and flag the record for manual review. Do not reward or penalize venue, citation count, novelty, writing style, or result magnitude. Return the decision, criterion, surface, centrality, paper type, evidence locations, exclusion code, quality flags, confidence, manual-review flag, and a one-sentence rationale.
\end{quote}
Both rounds used \texttt{DeepSeek-V4-Flash-0731}.

\paragraph{Consolidation and screening funnel.}
The source-level merge used lowercased DOI, version-stripped arXiv ID, source identifiers, and Unicode-normalized exact titles. DBLP records remained screenable from titles alone. 
The later work-level merge combined DOI and arXiv matches with fuzzy-title, abstract, and manual comparison. Duplicate or superseded versions are represented by one canonical record.

Table~\ref{tab:screening_funnel} in Section~\ref{ssec:search_protocol} summarizes the screening funnel.

\paragraph{Expert selection and snowballing.}
Expert review selected a work as \emph{representative} when it added a distinct mechanism, threat setting, control, evaluation resource, modality, domain, or empirical finding needed to cover the eligible landscape. \emph{Technically substantive} required a concrete method, benchmark, or measurement and enough full-text evidence for SLOT coding; venue prestige was not a criterion. 
Backward and forward citation snowballing then added BIPIA~\cite{yi2025benchmarking}, MINJA~\cite{dong2026memoryinjectionattacksllm}, and the canonical work by Greshake et al.~\cite{greshake2023not}.

\paragraph{Dates and released records.}
The earliest verified public date determined window eligibility. When metadata supplied only a year, \texttt{YYYY-01-01} records an explicit year-only fallback rather than a claimed publication day. 

The release documents the screening process and contains 555 records, including 135 selected works. 
It provides canonical identifiers and public paper or code links, screening-stage information and rationales, selected-paper summaries, and primary/secondary SLOT labels.

\end{document}